\documentclass[journal]{IEEEtran}
\usepackage{cite}
\usepackage{amsmath,amssymb,amsfonts}
\usepackage{algorithmic}
\usepackage{graphicx}
\usepackage{algorithm,algorithmic}
\usepackage{hyperref}
\hypersetup{hidelinks=true}
\usepackage{textcomp}
\usepackage{xurl}
\def\BibTeX{{\rm B\kern-.05em{\sc i\kern-.025em b}\kern-.08em
    T\kern-.1667em\lower.7ex\hbox{E}\kern-.125emX}}
\usepackage{atbegshi}
\AtBeginDocument{\AtBeginShipoutNext{\AtBeginShipoutDiscard}}
\begin{document}
\title{A Neuroimaging Simulation Framework for Developing and Evaluating Causal AI}
\author{Eryn Libert-Scott, Emma A.M. Stanley, Vibujithan Vigneshwaran, Matthias Wilms, Erik Y. Ohara, Nils D. Forkert}
\thanks{This work was supported in part by Alberta Innovates, in part by Natural Sciences and Engineering Research Council of Canada, in part by the University of Calgary’s Cumming School of Medicine, and in part by Canadian Neuroanalytics Scholars Program. (Corresponding author: Eryn Libert-Scott.) Eryn Libert-Scott (eryn.libertscott@ucalgary.ca) and Erik Y. Ohara (erik.ohara@ucalgary,ca) are with the Biomedical Engineering Graduate Program, University of Calgary, Calgary, AB, Canada. Eryn Libert-Scott, Vibujithan Vigneshwaran (vibujithan.vigneshwa@ucalgary.ca) , and Nils D. Forkert (nils.forkert@ucalgary.ca) are with the Department of Radiology, the Hotchkiss Brain Institute, the Alberta Children's Hospital Research Institute, and Department of Clinical Neuroscience, all at the University of Calgary, Calgary, Alberta, Canada. Emma A.M. Stanley (emma.stanley25@imperial.ac.uk) is with the Department of Computing, Imperial College London, London, U.K. Matthias Wilms (wilms@umich.edu) is with the Department of Radiology, University of Michigan, Ann Arbor, MI, USA }
\maketitle
\setcounter{page}{1}
\begin{abstract}
Causally linking disease-related factors to image-derived biomarkers provides a powerful pathway to understanding disease mechanisms. Despite growing interest in applying causal artificial intelligence (AI) approaches for this task, these methods still need to be adapted for complex medical images, and especially, neuroimaging. However, the lack of ground-truth data presents a barrier to development. To bridge this gap, we developed and tested a method for generating synthetic neuroimages, which adhere to a user-specified causal structure describing the non-image to image variable relationships, permitting the creation of ground-truth neuroimaging datasets. In the simulated T1-weighted magnetic resonance images, anatomical variability is modeled by sampling from a subspace estimated from real data and deforming a template image to create unique simulated subjects. Causal relationships are encoded via precise volumetric changes of any region-of-interest without unwanted global artifacts. We achieved relative volume errors of 0.3-2.66\% for the targeted regions-of-interest and demonstrate their statistically significant causal relationships, while maintaining mean absolute errors for non-target brain regions between 0.034-0.397ml. An initial evaluation of causal discovery methods exposes their limited ability to suppress spurious connections, highlighting the need for image-appropriate methods. Our framework is the first to enable the generation of realistic synthetic 3D neuroimages with explicit causal control that can serve as the missing ground-truth data necessary for the objective benchmarking and development of causal AI methods.
\end{abstract}


\section{Introduction}
\label{sec:introduction}
\IEEEPARstart{I}{nvestigating} and understanding cause-and-effect relationships is fundamental to scientific inquiry and essential for developing effective treatments for nearly all diseases. In medicine, randomized control trials (RCTs) remain the gold standard for identifying and validating causal relationships. However, RCTs are often expensive, time consuming, and sometimes even unethical or infeasible to conduct in humans. Furthermore, pre-clinical in-vitro and/or animal studies often do not always translate well to clinical outcomes \cite{lowenstein_uncertainty_2009}. These limitations motivate the development and application of causal artificial intelligence (AI) approaches that leverage the abundance of readily available healthcare data to approximate and supplement real-world experimentation.
 
Medical imaging has become one of the richest sources of information in the clinical routine and research, capturing many important aspects of the anatomy and organ of function. Therefore, imaging is frequently used to support clinical diagnosis, treatment planning, and investigation of diseases pathways. Although medical images often encode clinically relevant patterns, extracting those subtle indicators of pathologies reliably across patients can be challenging for the human eye \cite{mcleod_distinct_2025}. Within this context, previous work has demonstrated that machine learning (ML) techniques can support human experts by making use of these patterns to support clinical diagnosis and treatment decision \cite{topol_high-performance_2019} \cite{buch_artificial_2018}. 
Beyond prediction and pattern recognition in medicine, ML techniques have increasingly been applied to answer causal questions. ML-based causal analysis techniques are well established in several fields, such as economics, but have only recently gained traction in the medical imaging domain \cite{sanchez_causal_2022}, \cite{pearl_causality}, \cite{castro_causality_2020}. Within the causal AI landscape, two main fields exist, causal inference (CI) and causal discovery (CD). Briefly described, CI methods aim to estimate how interventions, such as treatment type, causally affect, for example, anatomical structures displayed in medical images and associated patient outcomes. CI techniques, like causal generative modeling, rely on pre-specified causal relationships often defined using domain knowledge \cite{komanduri2024identifiablecausalrepresentationscontrollable}\cite{kaddour_causal_2022}. However, in many cases there are no prior hypotheses to base the causal relationships on. Causal discovery (CD) methods offer a complementary data-driven alternative to define these relationships as they seek to identify the causal structure from observational data, among variables like risk factors and biomarkers. These causal structures are typically represented as directed acyclic graphs (DAGs) that are described by structural causal models (SCM). Together CD and CI have the potential to approximate the insights obtained from RCTs.

For many clinical applications the interest in applying CD and CI methods to medical images stems from the ability to better understand how clinical variables and/or demographics influence the physical manifestation of biomarkers. Given increasing number of people suffering from neurological conditions, neuroimaging may be of particular interest \cite{WHO}. For example, we can evaluate how age affects the presence of atrophy in the brain, or what differences may occur between male and females in buildup of amyloid-beta in the case of Alzheimer’s disease pathology. However, medical images, such as 3-dimensional brain magnetic resonance images (MRIs) are high-dimensional and spatially structured, with clinically relevant information encoded in both local and global regions with interconnected non-linear patterns. While conducting CD or CI on the entire image offers the ability to identify these relationships freely from the data, applying causal methods directly to full neuroimaging data remains challenging, especially for CD. Although recent CD methods, such as Non-combinatorial Optimization via Trace Exponential and Augmented lagRangeian for Structure learning (NOTEARS) \cite{notears} or Directed Acyclic Graph – Graph Neural Network (DAG-GNN) \cite{yu_dag-gnn_2019}, better handle higher-dimensional and non-linear data \cite{bello2023dagmalearningdagsmmatrices}, these methods still require tabularization or feature extraction, and cannot be applied directly to images. 

A fundamental obstacle in the development and evaluation of causal methods for neuroimaging is the lack of relevant ground-truth data. In real neuroimaging datasets, true causal structures are typically only partially observable. For example, by connecting clinical variables to imaging biomarkers, we are implicitly encompassing all the underlying biological processes that give rise to the biomarker manifestation which may be unknowable in practice. In a complete causal graph, we would also have to account for non-biological factors influencing the generation of the image itself, such as the scanner type. Considering the multitude of potential variables needing to be accounted for, it is most likely impossible to have access to a complete causal graph of real neuroimaging data. Therefore, it is problematic to evaluate the feasibility of causal methods on real-world neuroimage data, since there is generally no objective reference that the recovered graphs can be compared against.

In general, CD research to date has relied heavily on synthetic and semi-synthetic datasets for benchmarking and validation \cite{zanga_survey_2022}. Fully synthetic datasets are typically created by randomly generating DAGs and simulating data for the variables of interest accordingly. While this can provide complete causal ground truth data, this approach often fails to capture the complexity of real-world data. On the other hand, semi-synthetic data, like Tubingen cause-effect pairs \cite{causeeffect2026} and simulators like CausalWorld \cite{ahmed_causalworld_2020}, aim to improve realism while retaining partial controllability over the causal structure. However, the former is limited to bivariate relationships while CausalWorld, like many other semi-synthetic data generation methods, is highly domain specific. A small number of real-world datasets, primarily drawn from molecular biology, such as Sachs \cite{sachs_causal_2005}, Klein \cite{klein_droplet_2015}, and perturb-seq \cite{dixit_perturb-seq_2016}, have also been used to evaluate CD methods. While valuable, none of these datasets capture the unique challenges posed by complex neuroimages \cite{pezoulas_synthetic_2024}. Without datasets that resemble realistic neuroimages and include a known ground truth linking images and non-image variables, it is difficult to assess whether causal methods behave correctly. Enabling the generation of such data may help to advance CD and CI in medical imaging.

Synthetic neuroimages have been explored for tasks, such as data augmentation and modality translation, leveraging generative adversarial networks (GANs), variational autoencoders (VAEs), and diffusion-based models to reproduce realistic samples \cite{pezoulas_synthetic_2024}. While achieving realism from an appearance perspective, they do not enforce any known causal structure. More recently, generative modeling techniques have expanded to incorporate causality into the image generation processes, like CausalVAE \cite{yang2023causalvaestructuredcausaldisentanglement}, CausalGAN \cite{kocaoglu_causalgan_2017}, or MACAW \cite{vigneshwaran_macaw_2024}. These methods commonly enforce a predefined causal structure over learned latent image representations. However, regardless of an enforced causal structure, the data-driven latent variables that are learned do not necessarily correspond to generating factors in a direct, interpretable, or independent manner. Consequently, these forms of synthetic images are not ideally suited for evaluating causal AI methods. To ensure a fair and accurate evaluation of causal AI methods, a reproducible ground-truth dataset that enables explicit, interpretable control over the causal changes which are directly enforced on the images, is necessary. To the best of our knowledge, such a dataset for medical images does not currently exist.

Therefore, in this work, we introduce Simulated CAusal Representations in synthetic medical images (SCAR), a framework for generating synthetic but realistic neuroimaging datasets with known causal structures connecting non-image variables and image content. Practically, SCAR builds on the Simulated Bias in Medical Images (SimBA) framework proposed by Stanley et al. \cite{greenspan_flexible_2023} \cite{stanley_towards_2024},  that was developed to enable systematic evaluation of bias in ML models and lately also counterfactual interference \cite{gee_synthetic_2026}. Briefly described, SimBA generates synthetic brain MR images with controllable morphological biases in the form of realistic deformations. However, these deformations do not correspond to systematic changes in volume, and no explicit causal structure is enforced. The proposed SCAR framework adds the ability to control region-of-interest volumes and allows user-defined SCMs to be embedded directly into the data generation process. This creates the opportunity to generate ground-truth datasets suited for systematic evaluation of causal ML methods, including causal discovery, causal generative modeling, and counterfactual inference. In this work, we specifically focus on evaluating causal discovery as an initial feasibility demonstration and illustrate one practical application of our tool. The contributions of this work are as follows:
\begin{enumerate}
    \item We develop and describe a novel framework for generating simulated neuroimaging data that enables a direct, deterministic control of local regional volumes to impose causal links between non-image and images, providing ground-truth causal structures for evaluating causal algorithms on multimodal data relationships.
    \item We present an initial evaluation of five commonly used causal discovery methods and assess their ability to recover known causal graphs from the synthetic data generated using the developed framework.
\end{enumerate}
The SCAR framework is publicly available at https://github.com/erynl-s/SCAR 

\section{Methods}
\label{sec:methods}
\subsection{Overview}
The aim of the Simulated CAusal Representations in synthetic medical images (SCAR) framework is to generate 3D synthetic T1-weighted magnetic resonance image (MRI) scans of brains with controllable, ground-truth causal structures linking non-image variables to the generated images.
The process starts with a population template image, which is used as the basis to apply anatomical variability to simulate unique subject anatomies. Next, a structural causal model (SCM) is used to define how non-imaging variables influence specific brain regions. Based on these ground truth causal relationships, controlled deformations are applied to volumes of interest (VOIs) of each unique subject image to achieve target volumetric changes. In this work, the modified images are then evaluated to explore how closely the change in volume aligns with the target volume as specified by the causal relationship followed by testing various causal discovery algorithms to assess their ability to recover the known causal graphs.
\subsection{Data Generation}
\noindent\emph{Template and Dataset}

The proposed synthetic data generation begins with a standard brain MRI template. As in SimBA \cite{greenspan_flexible_2023}, \cite{stanley_towards_2024}, we used the SRI24 T1-weighted MRI template \cite{rohlfing_sri24_2010} as the reference image and the IXI database \cite{ixi2023} of healthy subjects to derive and model realistic inter-subject variability (ISV).

\noindent\emph{Region of Interest definition}

Segmentation of the template image was performed using SynthSeg \cite{billot_synthseg_2023}, producing binary masks and posterior probability maps for 32 anatomical regions. These regions include the third and fourth ventricles, brain stem and ventral DC for unilateral structures, cerebrospinal fluid (CSF) and total intracranial volume, then for each hemisphere: accumbens, amygdala, caudate, cerebellum cortex, cerebellum white matter, cerebral cortex, cerebral white matter, hippocampus, inferior lateral ventricles, lateral ventricles, pallidum, putamen, and thalamus. Binary masks provide a hard boundary of the segmented regions using 0 if the voxel is not part of the region and 1 if it is. The binary mask is derived from posterior probability maps, which provide a continuous value per voxel indicating SynthSeg’s confidence that the voxel belongs to the specific region.  These masks define the ROIs that can be targeted for causal interventions. 

\begin{figure}
    \centering
    \includegraphics[width=1\linewidth]{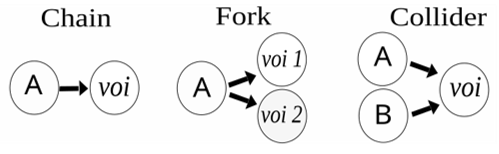}
    \caption{Directed graphs of three causal structures.}
    \label{fig. Fig 1}
\end{figure}

\noindent\emph{Inter-Subject Variability}

To create realistic subject-level variation, a subset of 50 subjects from the IXI database were, registered to the SRI24 template in a non-linear, diffeomorphic manner. The velocity fields from these registrations were analyzed using a principal component analysis (PCA), yielding a low-dimensional affine subspace of anatomical variation. Sampling velocity fields from this subspace and using them to deform the template brain MRI scan in a diffeomorphic i.e., topology-preserving manner, allows us to apply plausible local and global morphological variations, thereby simulating individual subjects with distinct, realistic brain morphology. This process follows the inter-subject variability (ISV) module introduced in SimBA  \cite{greenspan_flexible_2023}, \cite{stanley_towards_2024}.

\noindent\emph{Structural Causal Model}

SCAR introduces causality into the framework through a structural causal model (SCM) component that explicitly defines how non-imaging variables affect the morphology of specified, segmented brain VOIs.  Each non-imaging variable follows a user-defined statistical distribution that can be freely chosen, and the effect variables, derived from the user-defined SCM, determine the extent of volumetric change that needs to be applied to the VOI. In this work, we considered three canonical causal structures, depicted in Fig.~\ref{fig. Fig 1}. For the purposes of the preliminary analysis presented in the current work, we define arbitrary non-imaging variables A and B for the three causal structures. A is a scalar sampled from a normal distribution and B is a scalar sampled from a discrete distribution. In practical applications, these non-imaging variables can represent clinical or demographic factors, such as age, sex, education level, risk factors, etc. For the chain causal structure, the effect variable, d, for the targeted volume, voi, is defined as: 

\begin{equation}
d_{\text{voi}} = \alpha \times A + \epsilon
\label{eq:1}
\end{equation}

\noindent where \(\alpha\) is a linear scaling factor that modulates A’s effect on d and \(\epsilon\) is exogenous noise. Similarly, the fork defines the effect variables as in (\ref{eq:1}) for each voi: 
\begin{equation}
    \label{eq:2}
    \begin{aligned}
        d_{\text{voi}1} &= \alpha_1 \times A + \epsilon_1, \\
        d_{\text{voi}2} &= \alpha_2 \times A + \epsilon_2
    \end{aligned}
\end{equation}

\noindent In the case of the collider, the effect variable,d, depends on multiple variables: 
\begin{equation}
    d_{\text{voi}} = \alpha_1 \times A + \alpha_2 \times B +\epsilon
    \label{eq:3}
\end{equation}

\noindent\emph{Volume Control and Image Intervention}

The following steps are applied to each simulated subject created via the ISV protocol described above, after the full MRI scans with corresponding binary and posterior masks for all VOIs have been generated via deformation of the templates.
To achieve the target volume for each subject, we compute a new velocity field that expands or shrinks the spatial volume of the VOI in a series of steps. The goal is to adjust the target VOI volume in a controlled manner without introducing significant deformations outside the target VOI.

First, we construct a signed distance function (SDF). The SDF of the VOI binary mask provides each voxel’s distance from the initial VOI boundary. Let \(\chi\) be the image of the binary mask for the VOI and \(\Omega_{voi}\) be the volume of the VOI itself:
\begin{equation}
    \chi(x) \in \{0, 1\}, \quad \Omega_{\text{voi}} = \{ x : \chi(x) = 1 \}
    \label{eq:4}
\end{equation}

\noindent Then, the SDF \(\phi\), is defined by computing the distance between a voxel and the surface boundary of the VOI, \(\delta\Omega_{voi}\), where positive distances indicate being outside the VOI, negative inside the VOI and \(\phi(x)=0\) at \(\delta\Omega_{voi}\).

\begin{equation}
    \phi(x) =
    \begin{cases}
        -\operatorname{dist}\bigl(x, \partial\Omega\bigr), & x \in \Omega_{\text{voi}}, \\[4pt]
        \operatorname{dist}\bigl(x, \partial\Omega\bigr), & x \notin \Omega_{\text{voi}}.
    \end{cases}
    \label{eq:5}
\end{equation}		

\noindent The gradient of the SDF, \(\nabla\phi(x)\), yielding vectors normal to the surface, \(\delta\Omega_{voi}\), is used to define the velocity field,  \(v_{{\Omega}_{voi}}\), for the subject’s VOI. Importantly, the velocity field, \(v_{{\Omega}_{voi}}\), is restricted to a narrow band, B, around the boundary of the VOI. Constraining \(v_{{\Omega}_{voi}}\) to a small band is necessary to ensure that the deformation applied will make changes to the region’s volume based on boundary motion without global distortion, enabling a locally contained effect.

\begin{equation}
    v_{\Omega_V}(x) =
    \begin{cases}
        \dfrac{\nabla \phi(x)}{\lVert \nabla \phi(x) \rVert} \, \omega\bigl(\phi(x)\bigr), & x \in B, \\[6pt]
        0, & \text{otherwise}
    \end{cases}
        \label{eq:6}
\end{equation}

\noindent\text{where:}
\begin{equation}    
    B = \{ x : \lvert \phi(x) \rvert < b \}, \; b > 0.
    \label{eq:7}
\end{equation}

To ensure smooth local transitions and avoid sharp boundaries, a Gaussian fall-off, \(\omega(\phi)\), is applied so that the field is at full strength at boundary \(\phi=0\), and decays smoothly toward zero in either direction as \(|\phi|\) approaches \(b\). 

\begin{equation}
    \omega(\phi) = exp(\dfrac{-\phi^2}{\alpha \cdot {\sigma^2}}), |\phi|\leq b
    \label{eq:8}
\end{equation}

While the binary mask is used to define the velocity field, which is then used to impose changes on the simulated subject’s MRI scan, the posterior probability channel is used to calculate the baseline volume, \(V_{o}\), as the sum of the voxel probability values corresponding to that VOI. The posterior mask is used in the volume calculation instead of the binary mask to account for partial volume effects, thereby capturing a more representative volume. The majority of the volume contribution from the voxels included in the posterior mask and not in the binary mask is likely captured within the small band of the velocity field. The target volume \(V_t\) for the specified VOI is defined as:
\begin{equation}
    V_t^{voi}  = d_{voi} \times V_o^{voi}
    \label{eq:9}
\end{equation}

\noindent where \(d_{voi}\), is the effect variable as calculated as in (1), (2) or (3). We solve for a single scaling factor s that adjusts the magnitude of the velocity field, \(v_{{\Omega}_{voi}}\), such that the VOI volume matches the desired target volume, \(V_t^{voi}\). We formulate this as a root finding problem:
\begin{equation}
    f(s) = V_r (s) - V_t^{voi}  = 0
    \label{eq:10}
\end{equation}

\noindent where \(V_{voi} (s\)) is the volume computed from the posterior VOI mask after application of a deformation parameterized by \(s \cdot v_{{\sigma}_{voi}}\). Since \(V_{voi} (s)\) has no closed-form derivative, we employ the secant method \cite{nocedal2006backmatter} to iteratively update \(s\):

\begin{equation}
    s_{s+1}=s_k -f(s) \dfrac{s_k-s_{k-1}}{f(s_k)-f(s_{k-1})}
    \label{eq:11}
\end{equation}

For each iteration, \(V_{voi} (s\)) and \(f(s)\) are updated by applying the deformation \(s_{k+1} \cdot v_{\Omega_{voi}}\) directly to the posterior mask rather than the MRI scan, preventing additional processing steps. Once the optimal s is found, the final deformation is applied to the subject’s MRI scan.
\emph{Evaluation}
To evaluate the accuracy of the volume intervention, we compute and report on three quantitative measures, namely the mean absolute error (MAE), non-target MAE, and relative error. MAE measures how closely the achieved volume matches the target volume across subjects. It is calculated as the absolute error between the original volume of the subject’s target region and the volume post intervention, averaged across 1000 simulated subjects each with unique brain morphology.

\begin{equation}
   MAE=  \dfrac{1}{n} \sum_{i=1}^n|\hat{V}_{i,r_t}-V_{i,r_t}|
   \label{eq:12}
\end{equation}

\noindent where \(\hat{V}\) the intervened volume, \(V\) is the original volume, \(n\) is the number of subjects, and \(r_t\) is the intervened region. 
Non-target MAE quantifies the unintended changes to non-target brain regions. It is calculated as the absolute error between the original volume and the post-intervention volume of all non-intervened regions, averaged across subjects. While some change outside the target VOI must occur, to conserve the total volume, these changes should be limited to the neighboring regions and kept to a minimum overall.

\begin{equation}
    \text{Non-target MAE} =   \dfrac{1}{n} \sum_{i=1}^n\sum_{r \in R/r_t}{|\hat{V}_{i,R}-V_{i,R}}| 
    \label{eq:13}
\end{equation}

\noindent where \(R\) is the set of all regions, excluding only the target region. All volumes are extracted using SynthSeg as described above \cite{billot_synthseg_2023}.

Finally, the relative error is calculated as the absolute error of the target region volume, normalized for the size of the region. Although MAE reflects the discrepancy in absolute units, the relative error expresses this discrepancy as a percentage, enabling comparison across regions of varying size.

\begin{equation}
    \text{Relative Error} =  \dfrac{1}{n} \sum_{i=1}^n|\frac{{V}_{i,r_t}-V_{i,r_t}}{V_{i,R_t}}|
    \label{eq:14}
\end{equation}

To confirm that the simulated data followed the imposed causal relationships, we evaluated the relationship between the non-imaging variable, A, and the regional volumes using a linear mixed-effects model. The model included A, region, and their interaction as fixed effects, with a random intercept for subject to account for the multiple regional measurements contributed by each individual brain. The brain region was parameterized with the target region as the reference category, allowing direct comparison of each region’s slope with respect to A against that of the target region. This approach provided a quantitative assessment of whether the target region demonstrated the strongest relationship with A. A complementary visualization of the trajectory for each region against A was used to qualitatively confirm the pattern.

\subsection{Causal Discovery}

Existing causal discovery algorithms fall into various categories that differ with respect to their approach to identifying causal edges. To provide a representative evaluation across methodological paradigms, we selected algorithms spanning the major causal discovery approaches, including the Peter-Clark algorithm (PC) \cite{spirtes2000causation}, Greedy Equivalence Search (GES) \cite{maxwell2003optimal}, Linear Non-Gaussian Acyclic Model (LiNGAM) \cite{shimizu2006lingam}, Non-combinatorial Optimization via Trace Exponential and Augmented lagRangeian for Structure learning (NOTEARS) \cite{notears}, and Directed Acyclic Graphs via M-matrices for Acyclicity (DAGMA) \cite{bello2023dagmalearningdagsmmatrices} and evaluated their suitability for recovering the ground-truth relationships between non-imaging and imaging variables. For a more in depth review of causal discovery methods the reader is referred to \cite{zanga_survey_2022}, \cite{glymour_review_2019} and for causal concepts more generally to \cite{pearl_causality}.

Briefly described, constraint-based methods, such as PC, use conditional independence tests to determine the edges of the possible directed acyclic graph (DAG) configurations that can be removed from a graph of all variables that is initially fully connected, i.e., edges between all possible pairs of variables are present. Score-based methods, including GES, evaluate candidate graph structures by using a scoring function and select the structure with the most optimal score. GES does not score every possible DAG individually, but performs a search over equivalent graph structures, updating the current graph based on local changes that improve the score. LiNGAM infers directionality by exploiting the properties of non-Gaussian noise in the data-generating system i.e., causal model describing the data. Finally, optimization-based methods, exemplified by NOTEARS, are the newest class of causal discovery methods. Contrary to the other classes of causal discovery methods that are solving the DAG finding problem in discrete manner, searching over binary DAGs where edges are either present (1) or absent (0), optimization-based methods formulate the problem as continuous. This can be helpful because traditional binary approaches described above begin to fail with higher dimensions of variables due to the exponential increase in the size of the discrete search space as variables are added \cite{bickel_learning_1996}, \cite{chickering2004largesample}. Edges in optimization-based approaches, such as NOTEARS and DAGMA, are posed as weights between 0-1, thereby avoiding exponential discrete search with gradient-based optimization and scaling better to higher dimensions. 

Each causal discovery (CD) method tested in this work was implemented using publicly available Python packages NOTEARS, dagma, gcastle \cite{zhang_gcastle_2021} (PC and LiNGAM), and pgmpy \cite{ankan_pgmpy_2023} (GES). 

\noindent\emph{Experiments}

1,000 simulated subjects were generated for each of the three causal structure conditions as outlined above. Although our framework generates full 3D MRI scans, all CD methods described in literature so far are designed for structured, tabular variables and are not suited to work directly in the high-dimensional image space. Therefore, it was necessary to derive a set of meaningful tabular variables from the images to apply and test the methods appropriately. Therefore, all 3,000 simulated subjects’ MRIs were segmented using Synthseg \cite{billot_synthseg_2023} to extract volumes for 32 different brain regions (the same regions as described in the data generation process). The data input to the CD methods, therefore, consisted of non-imaging variables (A, B) and the extracted, tabular VOI volumes. The goal was to assess each algorithm’s ability to reconstruct the known ground-truth causal graph used to generate the imaging data using SCAR. To reduce the search space and isolate non-imaging-to-image causal effects, we constrained all methods so that VOI volumes could not serve as parents of other VOIs. For PC, LiNGAM, GES, and DAGMA, this constraint was incorporated using the prior-knowledge capabilities provided by each library. Since NOTEARS does not offer explicit prior constraints, we enforced this restriction through bound constraints on edge weights during optimization using the limited-memory Broyden–Fletcher–Goldfarb–Shanno algorithm \cite{nocedal2006largescale}, \cite{nocedal2006quasinewton}.

\noindent\emph{Evaluation Criteria}

The recovered graphs were compared to the imposed ground truth SCM using four commonly used metrics: 1) structural hamming distance (SHD), a measure of how close the estimated graph is to the ground truth graph, 2) true positive rate (TPR) to evaluate whether the existing relationships were correctly identified, 3) false positive rate (FPR) describing the number of non-existent edges that were incorrectly predicted relative to all non-edges, and 4) false discovery rate (FDR), which quantifies the proportion of predicted edges that were incorrect among all predicted edges. The PC and GES algorithms provide a binary map of edges, while the other methods provide a continuous value indicating the strength of the edges. Because PC and GES are highly sensitive to finite-sample variability, we applied bootstrapping 100 times to obtain edge frequency as a measure of stability. All methods were evaluated based on edges with weight or frequency values over 0.5 to reduce the likelihood of false positives and ensure only strong and/or stable connections are considered.

\section{Results}
\label{sec:results}
\subsection{Data Generation}

SCAR achieves highly accurate control of regional volumetric changes. Table~\ref{tab:target_volumes} summarizes intervention performance in the case of a chain causal structure for a representative sample of VOIs varying in size and shape, including the left lateral ventricle, left hippocampus, right amygdala, right putamen, and left cerebral cortex. The mean relative error for the target regions ranged from 0.30\% to 2.66\% , with the lateral ventricle producing the smallest error and the amygdala the greatest error. Corresponding target VOI MAE values remained as low as 0.03 ml for the left ventricle, and 3.886 ml for the largest region, the left cortex, indicating consistent adherence to the specified volumetric targets.

\begin{table*}[t]
\centering
\caption{Errors for the representative sample of volumes of interest with varying size and shape; left lateral ventricle, left hippocampus, right amygdala, right putamen, left cortex.}
\label{tab:target_volumes}

\resizebox{\textwidth}{!}{%
\begin{tabular}{|l|c|c|c|c|c|}
\hline
\textbf{Target Region} & 
\textbf{Mean target volume (ml)} &
\textbf{Mean achieved target volume (ml)} &
\textbf{Mean relative error (\%)} &
\textbf{Target ROI MAE (ml)} &
\textbf{Non-target MAE (ml)} \\
\hline
Left lateral ventricle & 10.778 & 10.803 & 0.30 & 0.030 & 0.051 \\
\hline
Left hippocampus       & 4.445  & 4.443  & 1.83 & 0.081 & 0.036 \\
\hline
Right amygdala         & 1.707  & 1.705  & 2.66 & 0.045 & 0.034 \\
\hline
Right Putamen          & 4.465  & 4.467  & 1.33 & 0.058 & 0.037 \\
\hline
Left cerebral cortex   & 230.587 & 230.380 & 1.70 & 3.886 & 0.397 \\
\hline
\end{tabular}
}
\end{table*}

\begin{table*}[t]
\centering
\caption{Linear mixed-effects model slope results for target region left lateral ventricle (reference) compared to all other regions (truncated).}
\label{tab:lme_slopes}

\resizebox{\textwidth}{!}{%
\begin{tabular}{|l|c|c|c|c|c|}
\hline
\textbf{Region} &
\textbf{Slope estimate} &
\textbf{Standard error} &
\textbf{p-value} &
\textbf{95\% CI (lower)} &
\textbf{95\% CI (upper)} \\
\hline
left lateral ventricle (reference) & 105.28 & 7.19 & 1.60e-48 & 91.18 & 119.37 \\
\hline
3rd ventricle & -104.94 & 9.93 & 4.37e-26 & -124.40 & -85.47 \\
\hline
4th ventricle & -106.21 & 9.93 & 1.11e-26 & -125.67 & -86.74 \\
\hline
brainstem & -109.14 & 9.93 & 4.39e-28 & -128.61 & -89.67 \\
\hline
cerebrospinal fluid & -96.58 & 9.93 & 2.40e-22 & -116.05 & -77.12 \\
\hline
Group Var & 0.05 & 0.00 & 1.02e-41 & 0.04 & 0.06 \\
\hline
\end{tabular}
}
\end{table*}

\begin{figure}
    \centering
    \includegraphics[width=1\linewidth]{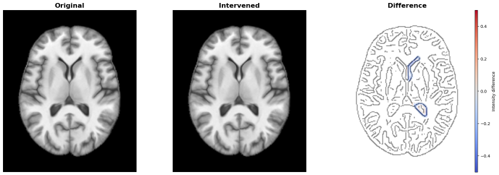}
    \caption{Original, intervened, and difference map images for left lateral ventricle on a subject.}
    \label{fig:2}
\end{figure}

Importantly, non-target MAE values were uniformly smaller than target MAE, confirming that unintended deformation beyond the intervention site was minimal. While some changes outside the target VOI are expected and required to occur to preserve the overall intracranial volume, the low values indicate the overall change was modest and limited to neighboring regions. Visual inspection of the difference maps shown in Fig.~\ref{fig:2} further revealed that the morphological changes are spatially localized to VOI boundaries without introducing artifacts or global distortions.

\noindent\emph{Validation of Causal Structure}

\begin{figure}
    \centering
    \includegraphics[width=1\linewidth]{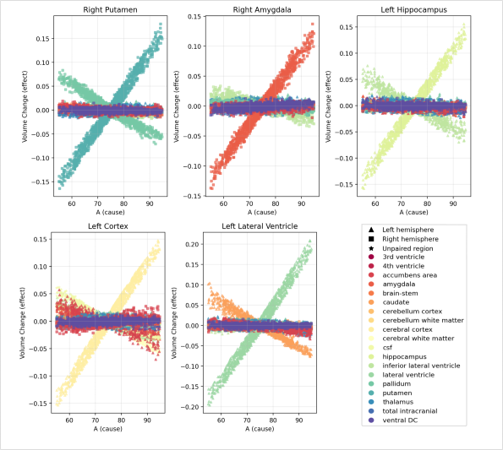}
    \caption{Volume change vs. causal variable across various target VOIs for the chain causal structure scenario.}
    \label{fig:3}
\end{figure}

Table~\ref{tab:lme_slopes} shows the linear mixed-effect model results for the target region of the left lateral ventricle, which indicate that the simulated data reflect the predefined structural causal models. For the full version of Table~\ref{tab:lme_slopes} please see the supplementary material. The slope of the left lateral ventricle with respect to the non-image cause variable A was found to be significantly different from all other regions as indicated by all p-values less than 0.001. Additionally, Fig.~\ref{fig:3} visually depicts the volume changes occurring for five different target regions with a single cause variable. As expected, only VOIs explicitly assigned as causal children displayed strong linear associations with their corresponding variables. In contrast, non-target regions exhibited negligible changes with the exception of the direct neighbors of the target regions, which is necessary and expected, to maintain brain topology. For those regions, the compensatory changes are consistent with intracranial volume conservation.

\subsection{Evaluation of Causal Discovery Methods}

\begin{figure}
    \centering
    \includegraphics[width=1\linewidth]{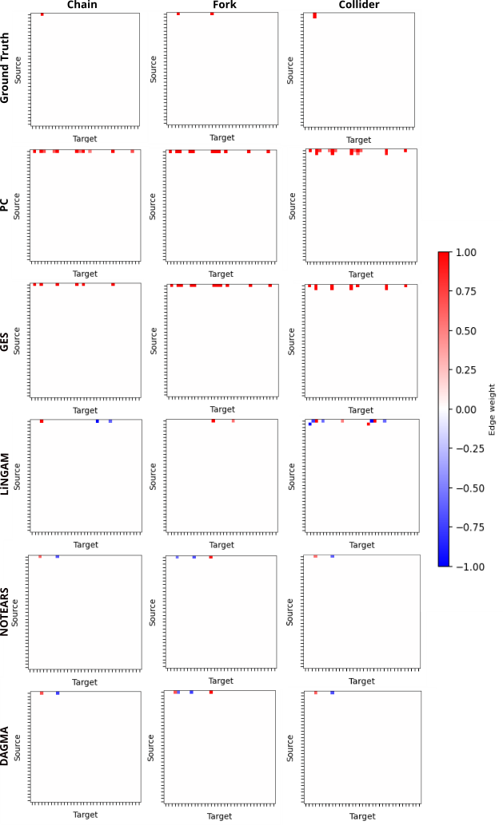}
    \caption{Adjacency matrices for the three ground-truth structures, chain(left), fork (middle), and collier (right) and the five causal discovery methods. Rows and columns in each matrix represent source (cause) and target (effect) variables respectively. Color represents the presence of an edge in the directed acyclic graph.}
    \label{fig:4}
\end{figure}

Five causal discovery algorithms (NOTEARS, DAGMA, PC, GES, and LiNGAM) were evaluated using the tabulated regional brain volumes extracted from segmentations of the SCAR-generated MRI datasets on three ground-truth causal structures: chain, fork, and collider. The resulting DAGs for each ground-truth scenario and each method are displayed as adjacency matrices in Fig.~\ref{fig:4}. Quantitative performance metrics are summarized in Table~\ref{tab:causal_metrics}.

\begin{table*}[t]
\centering
\caption{Causal discovery metrics including Structural Hamming Distance (SHD), True Positive Rate (TPR), False Positive Rate (FPR), and False Discovery Rate (FDR) for all four methods.}
\label{tab:causal_metrics}

\resizebox{\textwidth}{!}{%
\begin{tabular}{|l|ccc|ccc|ccc|ccc|}
\hline
\textbf{Method} &
\multicolumn{3}{c|}{\textbf{SHD} $\downarrow$} &
\multicolumn{3}{c|}{\textbf{TPR} $\uparrow$} &
\multicolumn{3}{c|}{\textbf{FPR} $\downarrow$} &
\multicolumn{3}{c|}{\textbf{FDR} $\downarrow$} \\
\hline
 & Chain & Fork & Collider & Chain & Fork & Collider & Chain & Fork & Collider & Chain & Fork & Collider \\
\hline
PC      & 10 & 9 & 15 & 1.0 & 1.0 & 1.0 & 0.322 & 0.300 & 0.234 & 0.909 & 0.818 & 0.882 \\
\hline
LINGAM  & 2 & 2 & 9 & 1.0 & 0.5 & 0.5 & 0.065 & 0.033 & 0.125 & 0.667 & 0.5 & 0.889 \\
\hline
GES     & 5 & 9 & 14 & 1.0 & 1.0 & 1.0 & 0.156 & 0.300 & 0.219 & 0.833 & 0.818 & 0.875 \\
\hline
NOTEARS & 1 & 1 & 2 & 1.0 & 1.0 & 0.5 & 0.032 & 0.033 & 0.016 & 0.5 & 0.333 & 0.5 \\
\hline
DAGMA   & 1 & 2 & 2 & 1.0 & 1.0 & 0.5 & 0.032 & 0.067 & 0.016 & 0.5 & 0.5 & 0.5 \\
\hline
\end{tabular}
}
\end{table*}

\noindent\emph{Chain}

All algorithms achieved a true positive rate (TPR) of 1.0, indicating that they consistently identified the true causal relationship between the non-imaging variable and the downstream brain regions. However, performance diverged in terms of graph accuracy and sparsity. NOTEARS and DAGMA achieved the lowest structural hamming distance (SHD) of 1 reflecting nearly perfect graph recovery with minimal extraneous edges corresponding to low false positive and false discovery rates (FPR = 0.032 and FDR = 0.5). LiNGAM performed next best (SHD = 2, FPR =0.065, FDR=0.667), whereas PC and GES introduced more false positives resulting in higher SHD (10 and 5, respectively), FPR (0.322and 0.156), and FDR values (0.909 and 0.833).

\noindent\emph{Fork}

For the fork ground-truth case, all algorithms except LiNGAM (TPR = 0.5), scored a TPR of 1.0 meaning both outgoing edges, from the single non-imaging parent to the two targeted brain regions, were consistently detected. NOTEARS performed best (SHD = 1), followed by LiNGAM and DAGMA (2), and PC and GES (9). Consequently, FPRs and FDRs were lowest for NOTEARS (FPR = 0.033, FDR = 0.333), LiNGAM (FPR=0.033, FDR=0.5), and DAGMA (FPR=0.067, FDR=0.5). PC and GES produced considerably denser graphs with FPR= 0.3 and FDR= 0.818, indicating inclusion of many more spurious edges. 

\noindent\emph{Collider}
    
The collider structure posed the greatest challenge across all methods. Only PC and GES maintained a TRP of 1.0, correctly detecting causal edges from each of the two non-image variables into the single targeted brain region. NOTEARS, DAGMA, and LiNGAM identified only one of the two incoming edges (TPR =0.5). However, NOTEARS and DAGMA were able to maintain lower SHD (2), FPR (0.016), and FDR (0.5) while PC, GES, and LiNGAM all had higher rates of SHD (14, 15, and 9, respectively), FPR (0.234, 0.125, and 0.219), and FDR (0.882, 0.889, and 0.875).

\begin{figure}
    \centering
    \includegraphics[width=1\linewidth]{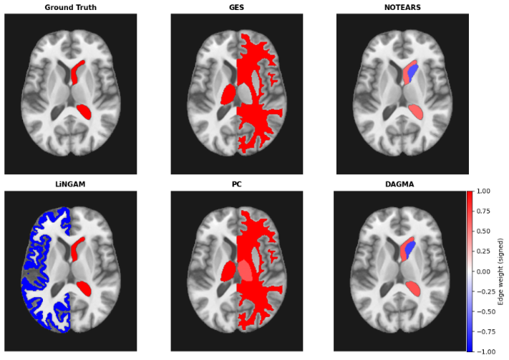}
    \caption{Visualization of causal graphs of the ground-truth (chain) and causal discovery methods, where highlighted regions represent edges identified in each respective directed acyclic graph.}
    \label{fig:5}
\end{figure}

The continuous optimization–based methods NOTEARS and DAGMA consistently produced the sparsest and most accurate graphs overall. Although PC and GES achieved perfect TPRs across all scenarios, they produced substantially denser graphs with many false positive edges. LiNGAM yielded intermediate sparsity but, like PC and GES, frequently included edges driven by spurious associations rather than true causal relationships. This behavior is illustrated in Fig.~\ref{fig:5} for the chain structure, where PC, GES, and LiNGAM identify causal edges into brain regions located in the contralateral hemisphere and distant from the true target region. In contrast, NOTEARS and DAGMA only introduce one additional edge, and it is anatomically adjacent to the target. Notably, the inferred causal effect for this neighboring region is opposite in sign to that of the target region, which is plausibly attributable to volumetric conservation effects and, therefore, represents a more interpretable and biologically consistent edge inclusion.

\section{Discussion}
\label{sec:discussion}

This study introduces and evaluates SCAR, a novel framework for generating synthetic but realistic MRI brain scans with explicitly defined causal relationships between non-imaging variables and images. The framework is primarily intended as a methodological research tool for developing and evaluating CI and CD methods for neuroimaging and multi-modal medical data. SCAR extends existing synthetic neuroimaging data generation, such as those using GANs or VAEs, by enabling direct volumetric manipulation of specific anatomical brain regions according to a user-defined structural causal model. It offers intensive flexibility, allowing users to select the atlas, target VOIs, variable distributions, and causal structures as appropriate to their research questions. This flexibility supports a broad range of experimental designs without constraining analyses to a fixed parcellation or predefined model.

The results demonstrate that SCAR can precisely impose regional volumetric changes while maintaining anatomic plausibility and minimizing unintended deformations in non-target brain areas, beyond those expected for conservation of intracranial volume. Specifically, non-target MAE results were as low as 0.034ml for the small amygdala region and 0.397ml for the large cortex region. These contained effects serve as an important advantage over other causal generative techniques, such as CausalVAE \cite{yang2023causalvaestructuredcausaldisentanglement} and CausalGAN \cite{kocaoglu_causalgan_2017}, whose data-generating processes can introduce changes into the images unrelated to and far away from the desired anatomical change \cite{sun_evaluation_2025}.

User-defined control over explicit causal and effect variables is key strength of the SCAR framework. Causal generative models impose a causal structure on learned latent variables. As such, it is not certain that the corresponding SCM has been accurately encoded because it is implicitly parameterized during model training. In contrast, SCAR enables specification of SCMs that define anatomically meaningful variables and outline customizable ground-truth relationships that are reflected in the image space directly, not latent variables. Because of these strictly enforced causal structures, SCAR can be used to systematically quantify how well these causal generative models encode causal structures and how closely their generated images match the ground-truth. Additionally, the SCAR framework can generate matched factual–counterfactual image pairs by simulating the same subject under varying conditions. These ground-truth pairs allow quantitative testing of both population-based inference and counterfactual image generation. Since deviations from the truth can be quantified, whether a model isolates the correct causal relationship or exploits spurious correlations can be tested. This supports evaluations that are difficult to achieve when using real neuroimaging data. 

Furthermore, as demonstrated in our preliminary analysis, CD methods can be evaluated using SCAR in the context of neuroimaging data. The CD experiments explored in this work revealed consistent challenges across all evaluated ground-truth causal structures. Although all algorithms, except LiNGAM, were able to correctly identify the non-imaging-to-VOI causal edges in the chain and fork settings, overall graph recovery varied substantially. The continuous optimization-based methods, NOTEARS and DAGMA, produced the sparsest graphs and lower structural errors. In contrast, PC and GES inferred more dense graphs with numerous spurious edges, despite recovering the true parent–child relationships.

The differences in CD method performance are likely attributed to correlations among the imaging-derived variables. Regional brain volumes exist within a physically constrained system (the intracranial volume), where changes to one structure necessarily induce compensatory changes in others. This interdependence leads to widespread correlations among VOIs, which can confound conditional independence testing or discrete graph search. Even when forbidding VOI-to-VOI edges, methods performing discrete search still inferred numerous spurious edges, illustrating the difficulty of disentangling true causal effects from spuriously correlated morphological patterns. More generally, this finding suggests that traditional CD methods, developed for tabular data, struggle with the structural dependencies and high dimensionality inherent to neuroimaging data, even when using tabularized image-derived features. Uncovering these insights were enabled by SCAR’s ground-truth data, which also supports more rigorous analyses of these methods and their applications or pitfalls. For example, including multiple simultaneous interventions, changes to intervention magnitude, or different spatial locations, enables systematic exploration of how causal structures affect downstream discovery.
Additionally, needing to select tabular image-derived features for CD restricts the analysis. Medical images contain an abundance of information that may not be well represented tabularly. Analyzing causal relationships on images directly would enable more informative, novel, and flexible exploration. Given that SCAR produces complete images, not just tabularized data, it naturally provides the opportunity for the development and validation of new CD methods that operate on images directly and tackle the challenges discussed. This would not be possible as efficiently without the framework provided by SCAR. 

Overall, the framework provides an environment that supports benchmarking causal machine learning methods under highly controlled, reproducible conditions. By turning causal adherence into a measurable property, SCAR enables development cycles in which causal AI methods for medical images can be iteratively stress-tested and refined against known failures.

Several limitations of the current framework should be acknowledged. First, SCAR prioritizes explicit causal control over biological completeness, and, while the generated images are realistic and anatomically plausible, they are not intended for precise disease modeling. Moreover, the velocity field band, b, was fixed across all regions to simplify implementation, but region-based optimization of b may further improve accuracy. The structural causal models used in this work were intentionally simple to establish a proof-of-concept. Extending SCAR to more complex causal graphs is easily achievable. Furthermore, adding the ability to model temporal dynamics would enhance its applicability. The non-image variable distributions employed in this work were arbitrarily chosen for illustrative purposes, they are not intended to replicate real demographic data. Finally, while the SCAR framework was developed on the case example of brain MRI scans, the methods presented in this work could be applied to other organs and/or image modalities. However, the interventional aspect i.e., changing regional brain volumes, may require some modification to be applicable in varying contexts.

\section{Conclusion}
\label{sec:conclusion}

The proposed SCAR framework presents the first approach for generating synthetic but realistic 3D medical brain images with explicitly defined causal relationships between non-imaging and imaging features. The application CD to extracted image features shows the limitations of these methods in the case of brain MRI scans and motivates the development of alternative approaches to causal analysis that better handle the unique dependencies present in medical images. By combining realism, interpretability, and causal control, SCAR lays the groundwork for the development of simulation tools aimed at understanding and improving causal reasoning in medical AI.

\bibliography{bibliography_scar}
\bibliographystyle{IEEEtran}
\end{document}